\documentclass[twocolumn,aps,10pt,aps,amsmath,amssymb,nofootinbib,superscriptaddress,showkeys,showpacs]{revtex4-1}
\usepackage{epsfig} 
\usepackage{amsmath} 
\usepackage{graphicx}
\usepackage{color}
\usepackage{float}
\usepackage{braket}

\begin{document} 

\title{Study of coherence and mixedness in meson and neutrino systems}

\author{Khushboo Dixit}
\email{dixit.1@iitj.ac.in}
\affiliation{Indian Institute of Technology Jodhpur, Jodhpur 342011, India}

\author{Javid Naikoo}
\email{naikoo.1@iitj.ac.in}
\affiliation{Indian Institute of Technology Jodhpur, Jodhpur 342011, India}

\author{Subhashish Banerjee}
\email{subhashish@iitj.ac.in}
\affiliation{Indian Institute of Technology Jodhpur, Jodhpur 342011, India}

\author{Ashutosh Kumar Alok}
\email{akalok@iitj.ac.in}
\affiliation{Indian Institute of Technology Jodhpur, Jodhpur 342011, India}

\date{\today} 

\begin{abstract}
We study the interplay between coherence and mixedness in meson and neutrino systems. The dynamics of the meson system is treated using the open quantum system approach taking into account the decaying nature of the system. Neutrino dynamics is studied in the context of three flavor oscillations within the framework of  a decoherence model recently used in the context of LSND (Liquid Scintillator Neutrino Detector) experiment.  For meson systems, the decoherence effect is negligible in the limit of zero CP violation. Interestingly, the average mixedness increases with time for about one lifetime of these particles. For neutrino system, in the  context of the model considered, the decoherence effect is maximum for neutrino energy around 30 MeV. Further, the effect of CP violating phase is found to decrease (increase) the  coherence in the upper $0< \delta < \pi$ (lower $\pi < \delta < 2\pi$) half plane. 
\end{abstract}

\maketitle 

%%%%%%%%%%%%%%%%%%%%%%%%%%%%%%%%%%%%%%%
\section{Introduction}
%%%%%%%%%%%%%%%%%%%%%%%%%%%%%%%%%%%%%%%
The quest for understanding the diverse nature of quantum correlations has lead in recent times to
a rethinking in terms of the most pristine adjective associated with quantumness, viz., quantum coherence. 
 Quantum coherence has been the subject of a lot of interest as a resource,  having implications to
implementation of technology \cite{marv,plenio,vaz} as well as in studies related to nonclassicality under the action of general open system effects \cite{SBhat}. The resource-driven
viewpoint is aimed at achieving  a better understanding of the classical-quantum
boundary,  in a mathematically rigorous fashion. 
Some of the well known measures of quantum coherence are those based on relative entropy \cite{plenio, Byrnes} and skew-information 
\cite{girolami}. Further, it has been recently shown that measures of entanglement can be put to use to 
understand quantum coherence \cite{adesso}. Recently a trade-off between mixedness and quantum coherence, from the perspective of {\it resource theory} \cite{resource}, inherent
in the system was proposed in the form of a complementarity relation \cite{uttam}.
These studies have mostly been focused on quantum optical systems from the perspective of quantum information. 

In this work we revisit these issues in the context of interesting sub-atomic systems, viz.  
the $K$ and $B$ mesons  that are copiously produced at the $\phi$  and $B$-factories, respectively and the neutrino system generated in reactors and accelerators.   These particles mainly aim at probing higher order corrections to the Standard Model (SM) of particle physics along with hunting for physics 
beyond the SM. The precision measurement at BaBar and Belle experiments confirmed the Cabibbo-Kobayasshi-Maskawa (CKM) paradigm of the SM.
The ongoing experiments at the Large Hadron Collider (LHC) and  Belle II experiments will now look for the signals of 
non standard $CP$ ($C$ stands for Charge Conjugation whereas $P$ stands for Parity) violating phases by very high precision measurements of various $B$ decays.
 On the other hand, the phenomena of neutrino oscillations is experimentally well established and ongoing and future experimental facilities in neutrino sector aim to resolve some unsolved problems such as observation of CP-violating phase in leptonic sector and the  mass hierarchy problem.  

These systems can also be used to study a number of fundamental issues in physics such as non-locality, entanglement 
\cite{BlasoneEPL2009,BlasonePRD2008,meson3,meson4,meson5,meson6,meson8,Banerjee:2014vga,Alok:2014gya,Banerjee,KD,Bramon:2005mg} as well as 
probing signatures of quantum gravity like background fluctuations \cite{Alok:2015iua}. Further, these systems are oscillating, unstable in nature and driven by weak interactions in contrast to the usual systems used in quantum information processing which 
are basically stable and governed by the electromagnetic interactions. Neutrinos are also weakly interacting and stable particles, showing oscillating behavior. Thus, the blend of all these makes these systems unique and interesting for the study of fundamental issues.

 In \cite{scully}, a subtle spilloff of the concept of wave-particle duality was brought out by the introduction of the concepts of 
quantum marking and erasure. This was quantified, in the language of interferometry, by the Englert-Greenberger-Yasin relation \cite{y,eg} which expressed 
a trade-off between the fringe visibility, a wave like nature, and path distinguishability, a particle like nature. This is also known as the quantum erasor and has
been investigated in the context of correlated neutral kaons \cite{bramonbeatrix}. Along with coherence, an important aspect of quantum dynamics is mixing which
quantifies the ability of the system of interest to retain its quantumness.

 Decaying systems are inherently open systems \cite{Banerjee:2014vga,Alok:2015iua}. Hence, it becomes relevant to understand the trade-off between
coherence and mixing in these decaying meson systems.
In this work we study the complementarity between coherence and mixing in these oscillating unstable open quantum systems driven by weak interactions.  We study how these are dependent on quantum gravity like background fluctuations, decay width and $CP$ violation, which is one of the important criterion to explain the present matter anti-matter asymmetry of the universe.

To undertake a similar study for neutrinos, we incorporate decoherence effects (possibly) due to quantum gravity like fluctuations in neutrino system, using the formalism of \cite{Farzan1,Farzan2}. This was constructed to explain the LSND (a short baseline accelerator neutrino experiment) data which could not be described by usual three flavor neutrino oscillation phenomena. We have also studied the CP-violating effects in this system.

 The paper is organized as follows. In Sec. (\ref{dynamics}), we discuss the dynamics of meson and neutrino system. The unstable meson system is describe in the language of  open quantum systems. In Sec. (\ref{measures}), we give a brief account of the measures of quantum coherence and mixedness.  Section (\ref{results}) is devoted to results and their discussion. We conclude in Sec. (\ref{conclusion}).

\section{System dynamics}\label{dynamics}
 In this section, we briefly discuss the dynamics of the meson and neutrino system. The former being a decaying system, is treated as an open quantum system. Neutrino dynamics is studied in the context of three flavor oscillations within the framework of  a decoherence model recently used in the context of LSND experiment.\par
\textit{Meson}. For the $B$ system, imagine the decay $\Upsilon\to b\bar b$ followed by hadronization into a $B\bar B$ pair. In the $\Upsilon$ rest frame, the mesons fly off in opposite directions (left and right, say); since the $\Upsilon$ is a spin-1 particle, they are in an antisymmetric spatial state. The same considerations apply to the $K$ system, with the $\Upsilon$ replaced by a $\phi$ meson.

The flavour-space wave function of the correlated $M\bar{M}$ meson systems ($M=K,\, B_d, \,B_s$) at the initial time $t=0$ is

\begin{equation}
\ket{\psi (0)} = \frac{1}{\sqrt{2}} \left[\ket{M \bar{M}} -\ket{\bar{M} M} \right], \label{flav}
\end{equation}

where the first (second) particle in each ket is the one flying off in the left (right) direction and $\ket{M}$ and $\ket{\bar{M}}$ are flavour eigenstates. As can be seen from \eqref{flav}, the initial state of the neutral meson system is a singlet (maximally entangled) state.  

The Hilbert space of a system of two correlated neutral mesons, as in \eqref{flav}, is  
\begin{eqnarray}
\mathcal{H}=(\mathcal{H}_{L}\oplus\mathcal{H}_0) \otimes (\mathcal{H}_{R}\oplus\mathcal{H}_0)\,,
\label{hilbert}
\end{eqnarray}
where $\mathcal{H}_{{L,R}}$ are the Hilbert spaces of the left-moving and right-moving decay products, each of which can be either a meson or an anti-meson, and $\mathcal{H}_0$ is that of the zero-particle (vacuum) state. Thus, the total Hilbert space is the tensor sum of a two-particle space, two one-particle spaces, and one  zero-particle state. The initial density matrix of the full system is 
\begin{equation}
\label{rhoh}
\rho_{\mathcal{H}}(0) = \ket{\psi (0)}\bra{\psi(0)}.
\end{equation}

The system, initially in the two-particle subspace, evolves in time into the full Hilbert space, eventually (after the decay of both particles) finding itself in the vacuum state. As can be appreciated from basic notions of quantum correlations such as entanglement, we need to project from the full Hilbert space $\mathcal{H}$ down to the two-particle sector $\mathcal{H}_{L}\otimes \mathcal{H}_{R}$. This is easily done once $\rho_{\mathcal{H}}(t)$ is written in the operator-sum representation \cite{caban,Banerjee:2014vga,Alok:2015iua}. The result is
\begin{equation}
\rho (t) = A
\left(\begin{array}{cccc} 
|r|^4 a_- & 0 & 0 & - \,r^2 a_-  \\ 0 & a_+ & -a_+  & 0 \\ 
0 &-a_+   & a_+ & 0 \\ - r^{*2} a_- & 0 & 0 & a_-
\end{array}\right),
\label{dm}
\end{equation}
where $a_\pm =  (e^{2\lambda t} \pm 1) (1\pm \delta_L)$, $A=(1-\delta_L)/4(e^{2\lambda t}-\delta^2_L)$, $\delta_L= 2 {\rm Re}(\epsilon)/(1+|\epsilon|^2)$ and $r=(1+\epsilon)/(1-\epsilon)$.  $\rho(t)$, which is written in the basis $\{\ket{MM},\ket{M\bar M},\ket{\bar{M}M},\ket{\bar{M}\bar{M}}\}$, is trace-preserving.  Here $\epsilon$ is a small $CP$-violating parameter \cite{Agashe:2014kda}. It is of order $\sim10^{-3}$  for $K$ mesons and $10^{-5}$ for $B_{d,s}$ mesons. $\lambda$ is the decoherence parameter, representing interaction between the one-particle system and its environment which could be ascribed to quantum gravity effects \cite{Hawking:1982dj,Ellis:1983jz,Banks:1983by,Ellis:1992eh,Ellis:1996fi,Ellis:2000sx,Mavromatos:2005bu,Mavromatos:2006yy,Mavromatos:2006yn,Feller:2016zuk}. Some quantum gravity models  are characterized by quantum fluctuations of space-time geometry, such as microscopic black holes, giving rise to quantum space-time foam backgrounds and hence may lead to decoherence. A stochastic fluctuation of point-like solitonic structure, known as D-particles, can constitute an environment for matter propagation \cite{Ellis:1996fi,Ellis:2000sx}. Further, if the ground state of quantum gravity consists of stochastically-fluctuating metrics, it can lead to decoherence \cite{Mavromatos:2006yy,Mavromatos:2006yn}. The decoherence in the mesonic systems can also be due to the detector background itself. Irrespective of the microscopic origin of the environment, its effect on the neutral meson systems, in our formalism, is modelled by a phenomenological parameter $\lambda$.

Diverse experimental techniques are used for testing decoherence, ranging from laboratory experiments  to astrophysical observations \cite{Benatti:2000ph,Ohlsson:2000mj,Benatti:2001tv,Ambrosino:2006vr,Fogli:2007tx,Mavromatos:2007hv,Oliveira:2014jsa,Bakhti:2015dca,Kerbikov:2015hug,Minar:2016rbm,Rastegin:2016xeb,Kerbikov:2015}. 
In the case of the $K$ meson system, its value has been obtained by the KLOE collaboration by studying the interference between the initially entangled kaons and the decay product in the channel $\phi \to K_S K_L \to \pi^+ \pi^- \pi^+ \pi^-$ \cite{Ambrosino:2006vr}. The value of $\lambda$ is at most $2.0 \times 10^8\, {\rm s^{-1}}$ at $\rm 3 \sigma$. Using existing Belle data on the time-dependent flavour asymmetry of semi-leptonic $B_d$ decays as given in Ref.~\cite{active}, an estimate on $\lambda$ for the $B_d$ system was obtained in \cite{Alok:2015iua}. At $\rm 3 \sigma$, the value of $\lambda$ is restricted to $0.012 \times 10^{12}\, {\rm s^{-1}}$ \cite{Alok:2015iua}. 
For $B_s$ mesons, to the best of our knowledge, there is no experimental information about $\lambda$. We consider it to be same as is in case of $B_d$ mesons. The mean life time for K, $B_d$ and $B_s$ mesons are $\tau_k = 1.7889 \times 10^{-10}$ sec, $\tau_{Bd} = 1.518 \times 10^{-12}$ sec and $\tau_{Bs} = 1.509 \times 10^{-12}$ sec, respectively. In case of K-mesons the CP-violating parameter is $|\epsilon| = 2.228 \times 10^{-3}$, $Re(\epsilon) = 1.596 \times 10^{-3}$. The dynamics of B-mesons is similar to that of K-meson with $\epsilon$ replaced by $\frac{p-q}{p+q}$. The decay widths are $\Gamma_k = 5.59 \times 10^{9} {\rm sec^{-1}}$, $\Gamma_{Bd} = 6.58 \times 10^{11} {\rm sec^{-1}}$ and $\Gamma_{Bs} = 6.645 \times 10^{11} {\rm sec^{-1}}$ \cite{Agashe:2014kda, hfag}.

Here the study of decaying systems is made using its evolution. Another approach would be to study, instead, the construction of effective
operators and study their evolution, instead of studying the state evolution, i.e., by invoking the Heisenberg picture. 
This approach was used to construct Bell inequality violations in neutral Kaon systems \cite{domenico}.\par

\textit{Neutrino}.
We study coherence in the context of three flavour neutrino oscillations. Three flavour states ($\ket{\nu_e}$, $\ket{\nu_{\mu}}$, $\ket{\nu_{\tau}}$) of neutrino mix via a $3\times3$ unitary (PMNS) matrix $U(\theta_{ij}, \delta)$, $i,j = 1, 2, 3$ ; $i<j$, where $\theta_{ij}$ are the mixing angles and $\delta$ is CP-violating phase, to form three mass eigenstates ($\ket{\nu_1}$, $\ket{\nu_2}$, $\ket{\nu_3}$). The mass eigenstates evolve as plane waves, i.e. $\nu_a(t) = e^{-iE_a t} ~\nu_a (0)$, $a=1,2,3$. A neutrino state $\ket{\Psi_\alpha (0)}$ at time $t=0$ evolves unitarily to  time t and can be written in flavor basis as
\begin{equation}\label{psi}
\ket{\Psi_{\alpha} (t)} = \sum_{\beta=e,\mu,\tau}^{}\zeta_{\alpha,\beta}(t) \ket{\nu_\beta}.
\end{equation}
Here $\zeta_{\alpha, \beta} = (UEU^{-1})_{\alpha, \beta}$, with $ E = diag [e^{-i E_1 t},e^{-i E_2 t},e^{-i E_3 t}]$. In this work we use the initial state as $\rho_{\mu} (0) = \ket{\nu_\mu}\bra{\nu_\mu}$, relevant to the LSND experiment and in the context of the decoherence model discussed in Sec. (\ref{measures}). The matrix $(UEU^{-1})$ contains, apart from the mixing angles $\theta_{ij}$, the mass square differences $\Delta m_{ij}^2 = m_j^2 - m_i^2$. The numerical values of these quantities used in this work are  $\theta_{12} = 33.48^o$, $\theta_{23} = 42.3^o$, $\theta_{13} = 8.5^o$, $\Delta m^2_{21} = 2.457 \times 10^{-3} {\rm eV^2}$, $\Delta m^2_{31} \approx \Delta m^2_{32} = 2.457 \times 10^{-3} {\rm eV^2}$ \cite{Garcia}.

\begin{figure*}[ht] 
	\centering
	\begin{tabular}{c}
		\includegraphics[width=150mm]{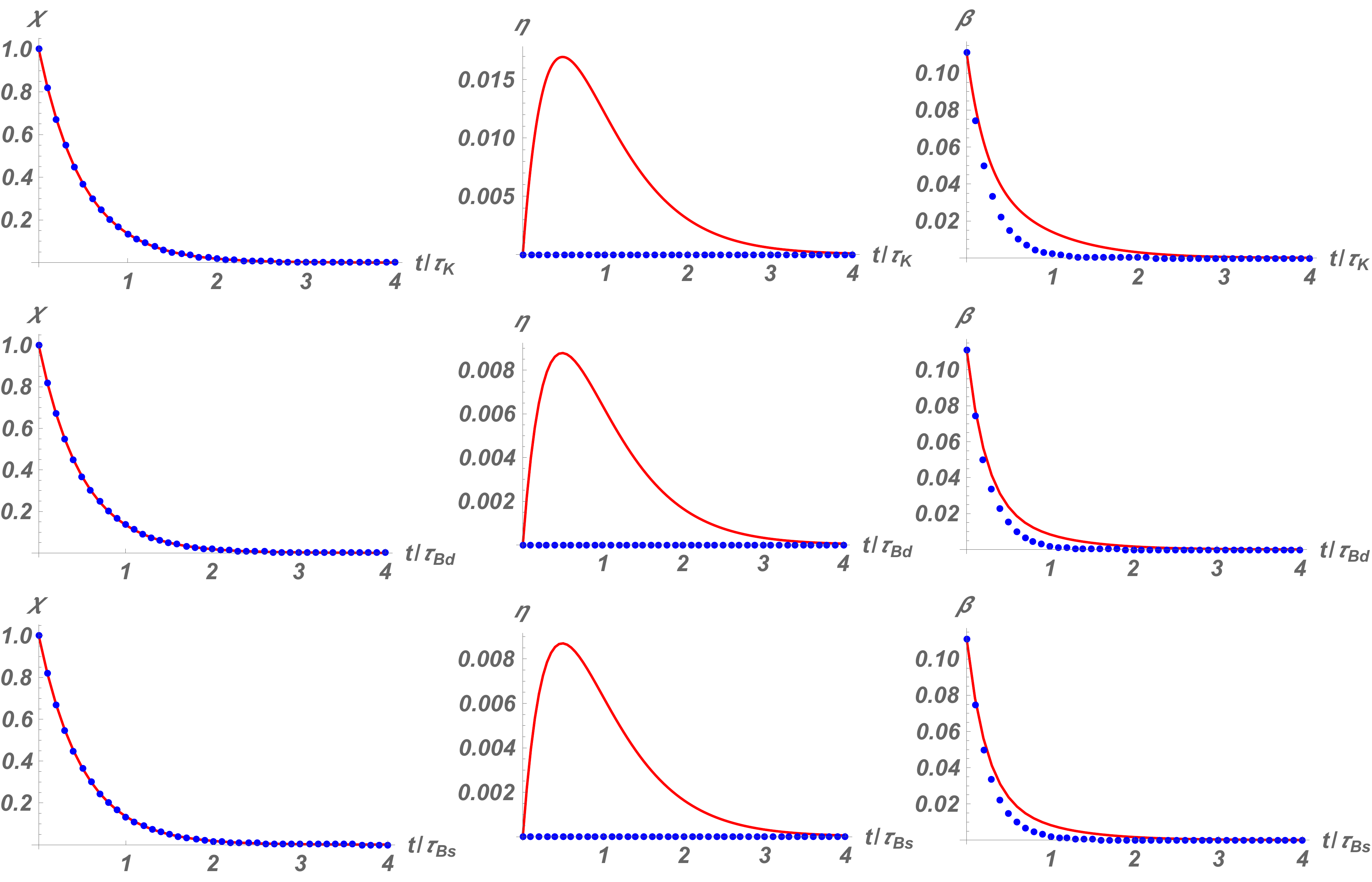}
	\end{tabular}
	\caption{(color conline) Meson system. Coherence parameter $\chi(\rho)$ (left), mixedness parameter $\eta(\rho)$ (middle) and complementarity parameter $\beta(\rho)$ (right) as function of the dimensionless quantity $t/\tau_{K (B_d, B_s)}$. Top, middle and bottom panels pertain to the case of K, $B_d$ and $B_s$ mesons, respectively. The solid (red) and dotted (blue) correspond to the case with and without decoherence, respectively.  In all the three cases, the average mixedness increases for about one lifetime of the particles.}
	\label{Joint}
\end{figure*}

\begin{figure*}[ht] 
	\centering
	\begin{tabular}{cc}
		\includegraphics[width=55mm]{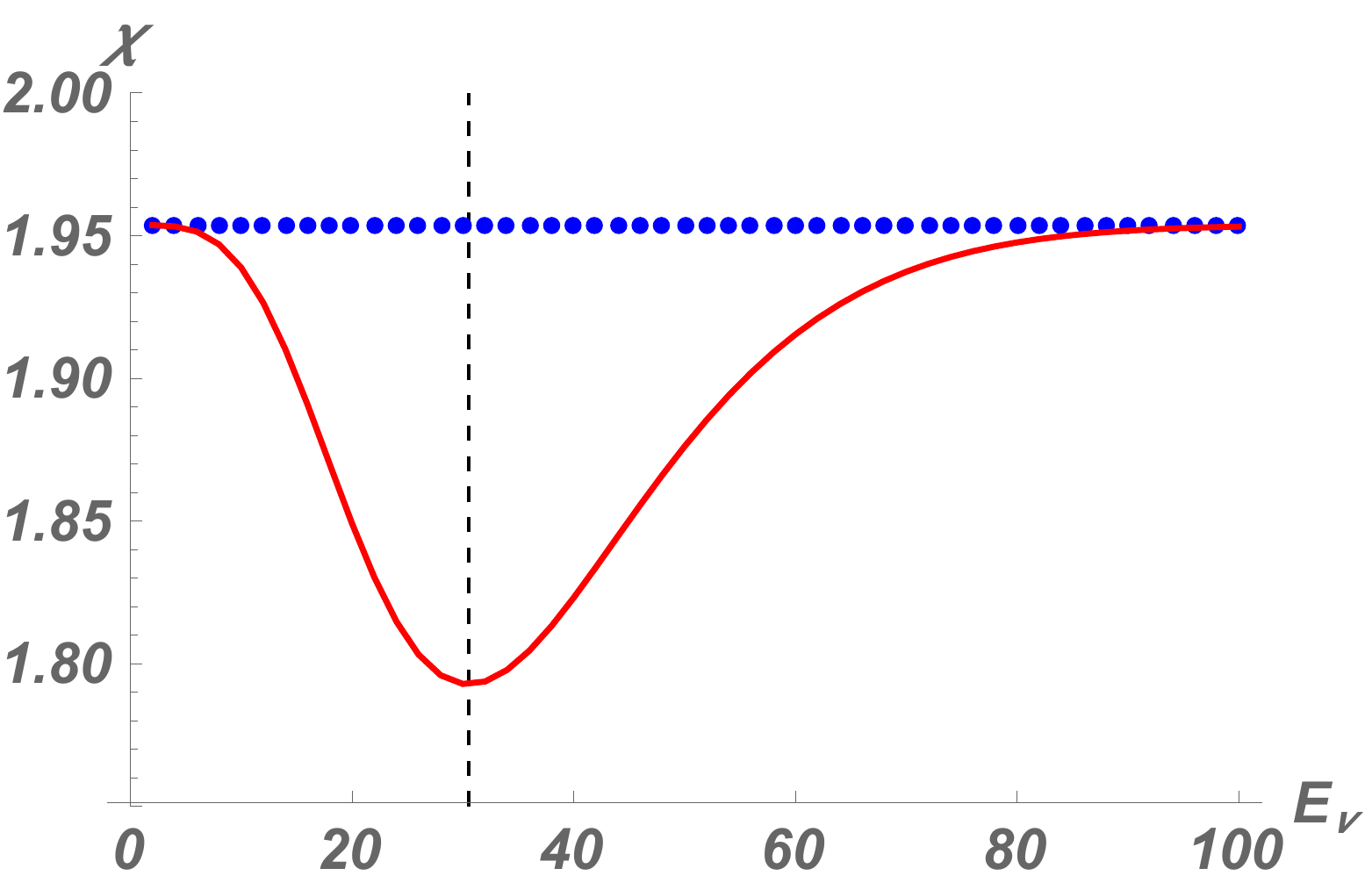}
		\includegraphics[width=55mm]{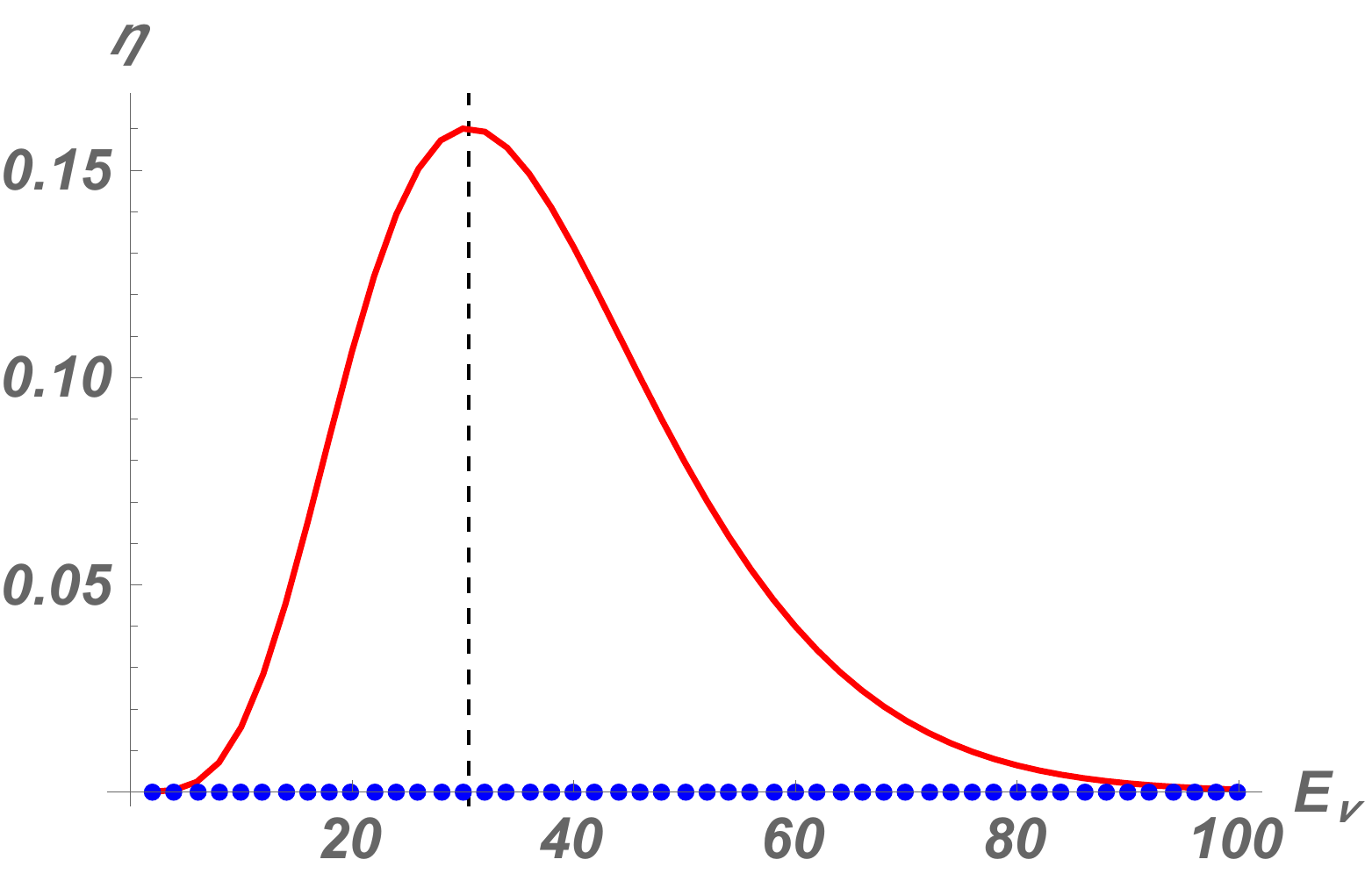}
		\includegraphics[width=55mm]{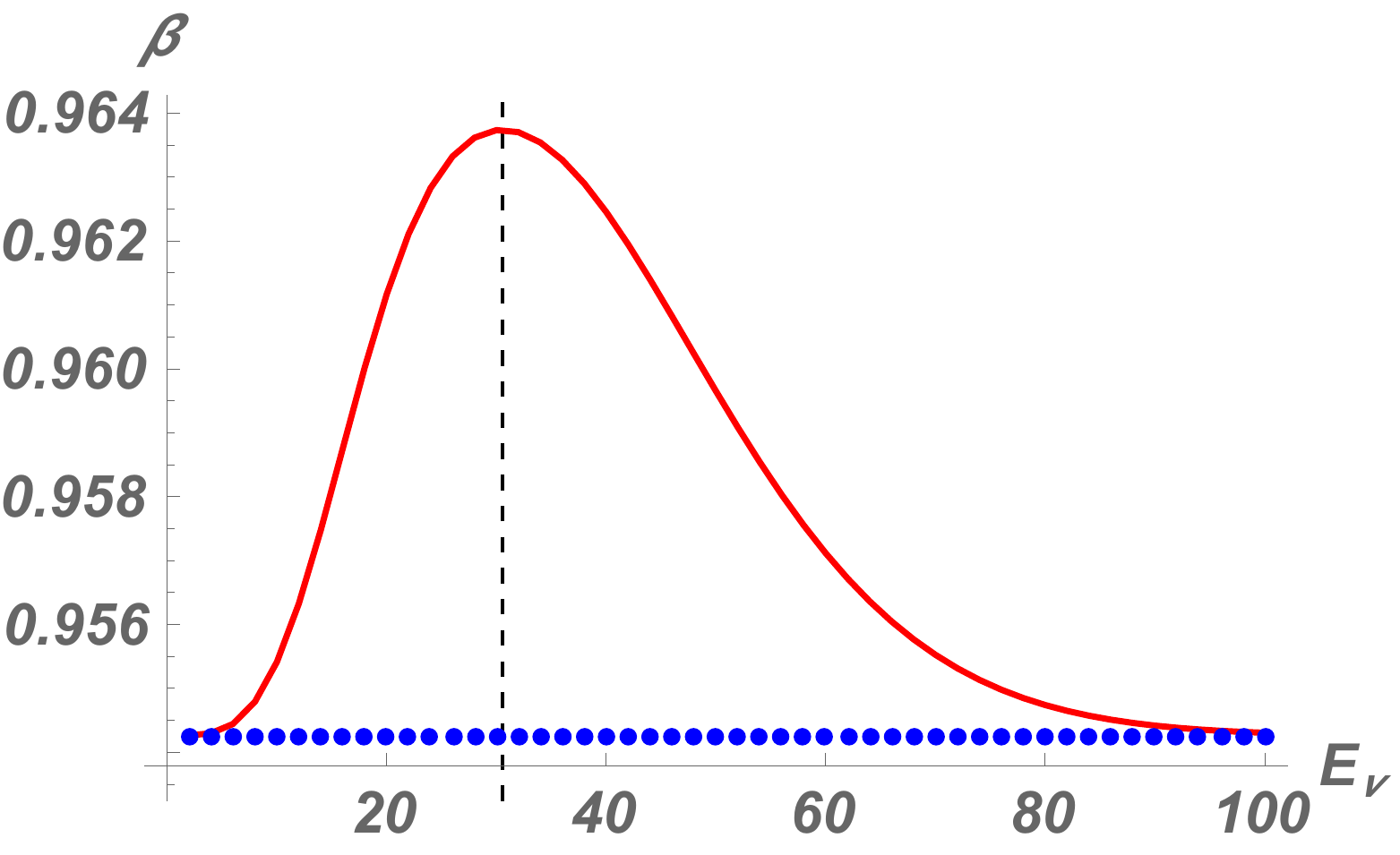}
	\end{tabular}
	\caption{(color conline) Neutrino system (LSND decoherence model). Coherence parameter $\chi(\rho)$ (left), mixedness parameter $\eta(\rho)$ (middle) and complementarity parameter $\beta(\rho)$ (right), plotted as a function of neutrino energy $E_{\nu}$ (MeV) with CP violating phase $\delta = 0$. The maximum value of decoherence parameter, defined in Eq. (\ref{gamma}), corresponds approximately to 30 MeV (vertical dashed  line). At this energy the coherence and the mixedness parameter attain their minimum and maximum values, respectively.}
	\label{CohMxd_NeuBD}
\end{figure*}

\begin{figure}[ht] 
	%\centering
	\includegraphics[width=70mm]{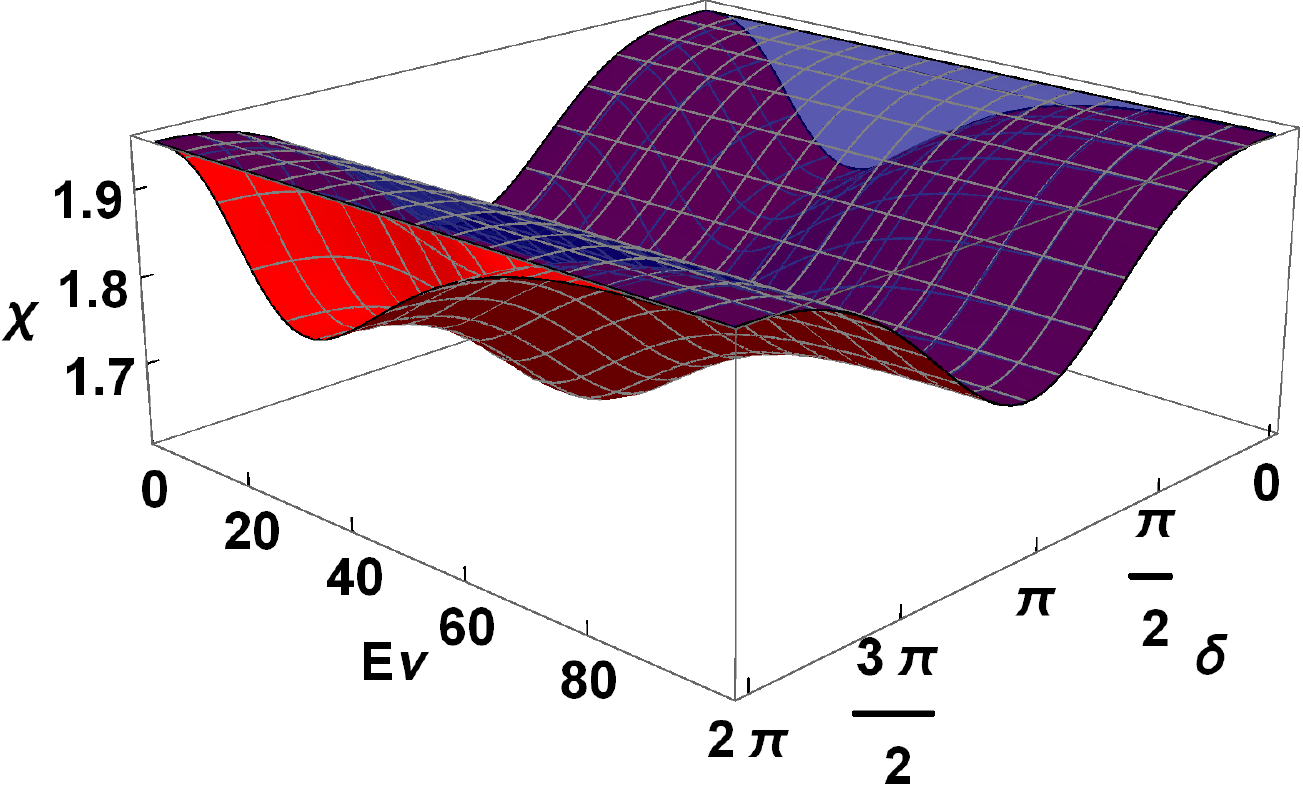}
	\caption{(color conline) Coherence parameter $\chi$ as defined in Eq. (\ref{chiBD}), for the state in Eq. (\ref{rhom}), is plotted with respect to the CP violating phase $\delta$ and the energy $E_\nu$ of the neutrino. The blue and red surfaces correspond to the cases with and without decoherence parameter, respectively. The minimum of the coherence parameter occurs for $E_\nu \approx 30 ~{\rm MeV}$ and $\delta = \pi$. }
	\label{CP}
\end{figure}

\section{Measures of coherence and mixedness}\label{measures}
Here we study the interplay between quantum coherence and mixedness inherent in the state (\ref{dm}). We use the definition of coherence $\chi(\rho)$ as given in \cite{Alok:2014gya}, in terms of the off-diagonal elements of the density matrix. The mixedness parameter denoted by $\eta(\rho)$ is given in terms of the normalized linear entropy \cite{peters}. The balance between coherence and mixedness was recently expressed in terms of a complementarity relation as \cite{uttam}
\begin{equation}
\beta(\rho) = \frac{\chi^2(\rho)}{(n-1)^2} + \eta(\rho)\leq 1. \label{cohmix}
\end{equation}
Here
\begin{align}
\chi(\rho) &= \sum_{i\neq j} |\rho_{ij}|,\label{chiBD} \\
\eta(\rho) &=  \frac{n}{n-1}(1-{\rm Tr}\,\rho^2) \label{etadef},
\end{align}
where $\rho_{ij}$  are the off-diagonal elements of the density matrix $\rho$ and  $n$ is the dimension of the system. For meson and neutrino systems described in the previous section, $n=4$ and $3$, respectively.\par
\textit{Coherence and Mixedness in meson system}. In order to take into account the effect of decay in the system under study, the measures given in Eqs. (\ref{chiBD}) and (\ref{etadef}) must be multiplied by the probability of survival of the pair of particles up to that time, $P_S(t)$, which can be shown to be 
\begin{equation}
P_S(t)= e^{-2\Gamma t}\frac{(1-\delta^2_L e^{-2\lambda t})}{1-\delta^2_L},
\label{pest}
\end{equation}
where $\Gamma$ is the meson decay width. For the $K$ meson system, $\Gamma=(\Gamma_S + \Gamma_L)/2$ where $\Gamma_S$ and $\Gamma_L$ are the decay widths of the short and the long neutral Kaon states, respectively.

For the state given in Eq. (\ref{dm}), we have
\begin{equation}
\chi(\rho)=2 P_S(t) A \left[a_+ + a_- \left|\left(\frac{\epsilon+1}{\epsilon-1}\right)^2\right|\right].
\label{chi}
\end{equation}
 Also, the mixedness parameter $\eta(\rho)$ is given by
 \begin{equation}
 \eta(\rho)=\frac{4}{3}P_S(t)\left[1-a_-^2 A^2 \left\{1+4\left(\frac{a_+}{a_-}\right)^2 +2\delta_\epsilon^2 +\delta_\epsilon^4\right\} \right]\,,
 \label{eta}
 \end{equation}
where
\begin{equation}
\delta_\epsilon=\frac{1+2{\rm Re}(\epsilon)+|\epsilon|^2}{1-2{\rm Re}(\epsilon)+|\epsilon|^2}\,.
\end{equation}
 The corresponding simplified expressions, neglecting the small CP violation, can be obtained from these expressions by setting $\epsilon$ equal to zero.\par
 
\textit{Coherence and mixedness in neutrino system.} In order to understand the interplay between coherence and mixedness for the  three flavor neutrino system, we need to take into account the decoherence effects due to environmental influences. For this purpose we follow the phenomenological approach of \cite{Farzan1} whose motivation was to explain the LSND signal via quantum-decoherence of the mass states leading to the damping of the interference terms in the oscillation probabilities. A possible source of this kind of effect might be quantum gravity. The decoherence effects when taken into account lead to the following  form of the density matrix in mass basis \cite{Farzan1}
	\begin{widetext}
	\begin{equation}
	\rho (t) = 
	\begin{pmatrix}
	\rho_{11}(0)                     &\rho_{12}(0) e^{-(\gamma_{12} - i \Delta_{12})t}      &\rho_{13}(0) e^{-(\gamma_{13} - i \Delta_{13})t}\\
	\rho_{21}(0) e^{-(\gamma_{21} - i \Delta_{21})t}                      &\rho_{22}(0)                        &\rho_{23}(0) e^{-(\gamma_{23} - i \Delta_{23})t}\\
	\rho_{31}(0) e^{-(\gamma_{31} - i \Delta_{31})t}    &\rho_{32}(0) e^{-(\gamma_{32} - i \Delta_{32})t}                           &\rho_{33}(0)
	\end{pmatrix}.\label{rhom} 	
	\end{equation}
	\end{widetext}
Here $\gamma_{ij}$ are the decoherence parameters and $\Delta_{ij} \approx \frac{\Delta m_{ij}^2}{2E_\nu}$, with $ \Delta m_{ij}^2$ and $E_\nu$ being the mass square difference and  the energy of the neutrino, respectively. Here the elements of the density matrix at time $t=0$ are given by $\rho_{ij} (0) = U_{\alpha i}^* U_{\alpha j}$, such that $U_{\alpha i}$ are the elements of the PMNS (Pontecorvo--Maki--Nakagawa--Sakata) matrix. 
%The state of the system can be written in flavor state basis as:
%	\begin{equation}
%	\rho_f (t) = U \rho(t) U^\dagger.
%	\end{equation} 
	The most economic scenario that describes all the LSND data leads to 
	\begin{equation}
	\gamma_{12} = 0  \quad {\rm and} ~\gamma \equiv \gamma_{13} = \gamma_{32}.
	\end{equation}
	In \cite{Farzan2} an exponential dependence of the decoherence parameters on neutrino energy was conjectured to be
	\begin{equation}\label{gamma}
	\gamma = \gamma_0 \bigg(\exp\bigg[-\bigg( \frac{E}{E_3}\bigg)^n\bigg] - \exp\bigg[-\bigg( \frac{E}{E_1}\bigg)^n\bigg]\bigg)^2.
	\end{equation}
	Best fit values of $E_3$ and $E_1$ are 55 MeV and 20 MeV, respectively with $n = 2$ and $\gamma = 0.01 ~{\rm m^{-1}}$. It turns out that $\gamma$ attains its maximum value at around 30 MeV; consequently, one expects  maximum decoherence at this energy.
%	The coherence measure given in Eq. (\ref{chi}) is a basis dependent quantity. As shown in Fig. (\ref{CohMxd_NeuBD}), in mass basis, coherence of the state decreases as $\gamma$ increases, as expected. However,  this feature is lost when when the state is transformed to flavor basis, Fig. (\ref{CohMxd_NeuBI}). To avoid this discrepancy we use a basis independent measure of coherence in terms of the entropy, recently introduced in \cite{Byrnes}:
%	\begin{equation}\label{chitilde}
%	\tilde{\chi} = \sqrt{S \bigg(\frac{\rho + \mathbf{I}/d}{2}\bigg) - \frac{S(\rho) + \log_2 d}{2}}.
%	\end{equation}
%	Here $d$ is dimension of the system and S($\rho$) represents the von-Neumann entropy of the state $\rho$, as
%	\begin{equation*}
%	S(\rho) = -Tr(\rho~ \log_2 \rho).
%	\end{equation*}

%\begin{figure}
%	\includegraphics[width=55mm]{Neu_BD_FB}
%	\includegraphics[width=55mm]{Neu_BI}
%	\caption{Coherence parameter  for neutrino system in \textit{flavor  basis} (top) as defined in Eq. (\ref{chiBD}) and by basis independent definition (bottom) given in Eq. (\ref{chitilde}). At approximately 30 MeV, the decoherence parameter $\gamma$ (Eq. (\ref{gamma})) becomes maximum and we expect a minimum value of the coherence parameter. The unexpected behavior in the former case is because of basis dependent structure of the model in Eq. (\ref{rhom}).}\label{CohMxd_NeuBI}
%\end{figure}

%%%%%%%%%%%%%%%%%%%%%%%%%%%%%%%%%%%%%%%
\section{Results and discussions}\label{results}
%%%%%%%%%%%%%%%%%%%%%%%%%%%%%%%%%%%%%%%

We now interpret the dynamics of meson decay and neutrino oscillations from the perspective of concepts underpinning the foundational aspects of quantum mechanics, such as the interplay between coherence and mixing.\par

\textit{Mesons:}~For meson-systems, coherence is a function of $\lambda$, $\epsilon$, $t$ and $\Gamma$. For $\lambda=0$, $\chi = e^{-2\,t\, \Gamma}$. Thus we see that, in the absence of decoherence, coherence depends only on $\Gamma$ and $t$ and not on $CP$ violation. In the absence of $CP$ violation in mixing, $\chi = e^{-2\,t\, \Gamma}$. Hence in the limit of neglecting $CP$ violation in mixing, coherence is independent of decoherence parameter.

Like coherence, mixedness is also function of $\lambda$, $\epsilon$, $t$ and $\Gamma$. $\eta(\rho)=0$ only if $\lambda=0$, i.e., the state (\ref{dm}) will become mixed only in the presence of quantum gravity like background fluctuations.  Also the maximum value of $\eta(\rho)$ cannot  approach 1, as can be seen from Fig.(\ref{Joint}). Hence the concept of maximally coherent mixed state, which is valid for states satisfying the equality in the complementarity equation will never happen in these decaying systems, as can be seen from the right most panels of Fig (\ref{Joint}). Apart from the violation of the well-known relation between non-locality and teleportation fidelity, as observed in \cite{Banerjee:2014vga}, this brings out another difference between the stable and decaying system. 
The effect of $CP$ violating parameter $\epsilon$ in mixedness is negligible.
It is obvious from the middle panel of Fig. (\ref{Joint}) that $\eta(\rho)$ increases with $t$ till about one life time of the mesons. After one life time, the average mixedness is seen to decrease with time. This is an artifact of the decaying nature of the system and can be attributed to the modulation by $P_S(t)$, Eq.~(\ref{pest}). 

The interplay between coherence and concurrence, $C(\rho)$, can be represented as
\begin{equation}
C(\rho) = \chi(\rho) \, e^{-2t \lambda}.
\label{coh-con}
\end{equation}
It can be seen from the above relation that coherence and concurrence are not synonyms with each other, a fact that is highlighted by the interest in coherence resource 
theory \cite{plenio}.
Further, the interplay between coherence, mixedness and nonlocality \footnote{Here we do not study an experimental test of Bell's inequality from local realism. Instead, we make use of an important result obtained in \cite{horo} which enables us to make quantitative statements about Bell inequality violations, indicative of nonlocality, just by making use of the parameters of the density operator describing the system.}, $M(\rho)$, is given by 

\begin{equation}
\eta(\rho) =\frac{4}{3} \left( \chi(\rho)- \frac{M(\rho)}{2}\right).
\label{mix-nl}
\end{equation}\\
Here $M(\rho)$ denotes non-locality modulated by the probability of
survival of the pair of particles up to the time of interest \cite{Banerjee:2014vga}. $M(\rho)>1$ implies non-locality \cite{horo}.
To keep the expression simple, here we have neglected the small $CP$ violating parameter $\epsilon$.
From the above equation, we get $M(\rho)  = 2 \chi(\rho)- \frac{3}{2}\eta(\rho)$.
Using the theoretical expressions for $\chi(\rho)$ and $\eta(\rho)$, given in Eqs. (\ref{chi}) and (\ref{eta}), respectively, along with the numerical inputs for $\Gamma$ and $\lambda$, 
we can reproduce the $M(\rho)$ plots as obtained in \cite{Banerjee:2014vga}. This brings out  the consistency of the present analysis.

\textit{Neutrinos:}~{In case of neutrinos, since we are restricted to use a specific model to incorporate decoherence effects, we used the phenomenological approach of  \cite{Farzan1,Farzan2} which was motivated to explain the LSND signal. This model considers the decoherence effect in mass basis. The decoherence parameter $\gamma$, given in Eq. (\ref{gamma}), has exponential dependence on neutrino-energy $E_{\nu}$ and attains maximum value at approximately 30 Mev. Consequently, the coherence parameter $\chi$ and the mixedness parameter $\eta$ attain their minimum and maximum values, respectively, at this energy. The complementarity relation also holds in case of neutrinos as can be seen in right panel of Fig. (\ref{CohMxd_NeuBD}).\par
 The coherence for the neutrino system is found to depend on the CP violating phase $\delta$. The effect of the CP violating phase $\delta$ is depicted in Fig. (\ref{CP}). It is clear that the  coherence parameter decreases (increases) in the upper half plane $0< \delta < \pi$ (lower half plane $\pi < \delta < 2\pi$), and attains its minimum value at $\delta = \pi$ and energy $E_\nu \approx 30~ {\rm MeV}$.
%	 Figure (\ref{CohMxd_NeuBI})  depicts the coherence parameter $\chi$ in flavor basis. The unexpected increase in $\chi$ with the increase in $\gamma$ is due to the basis dependent structure of the model considered. 
%	By using the recently introduced basis independent measure of coherence $\tilde{\chi}$ (Eq. (\ref{chitilde})), we find the expected behavior of the coherence parameter, that is, a decrease in coherence in presence of the decoherence effect and observed negligible effects of CP-violating phase on these measures.           

%%%%%%%%%%%%%%%%%%%%%%%%%%%%%%%%%%%%%%%
\section{Conclusions}\label{conclusion}
%%%%%%%%%%%%%%%%%%%%%%%%%%%%%%%%%%%%%%%

In this work we have studied the interplay between coherence and mixedness for the systems such as neutral mesons and neutrinos which are governed by weak interactions.
The meson systems are interesting as well as challenging due to them possessing both oscillatory and decaying nature while neutrino system is stable. We study the impact of decoherence and $CP$ violation.  For mesons, in the limit of neglecting $CP$ violation in mixing, it is observed that coherence is independent of the decoherence parameter. It is also shown that the concept of maximally coherent mixed state, which is valid for states satisfying the equality in the complementarity relation  will never happen for the correlated neutral meson system. 
An interesting feature that comes out is that for about one life time of these particles, the average mixedenss increases with time in consonance with our 
usual notion of mixedness. However, after this, the mixedness decreases with time, a behavior that can be attributed to the decaying nature of the system. Further, we also find that for these correlated meson systems, quantum coherence and entanglement are not synonyms with each other, a fact that has been bolstered by the interest in coherence resource theory.

For neutrino system, coherence and mixedness is studied in the context of the  decoherence model for LSND experiment. In this model, the decoherence parameter $\gamma$ is a function of the energy of neutrino $E_\nu$. The coherence parameter  decreases with the increase in $\gamma$ and attains its minimum value at $ E_\nu \approx 30~ {\rm MeV}$. Further, the coherence for the neutrino system is found to depend on the CP violating phase $\delta$. %such that it decreases (increases) in the upper half plane  (lower half plane). 

\end{document}